\documentclass[aps,reprint,superscriptaddress]{revtex4-1}

\usepackage{graphicx}
\usepackage{dcolumn}
\usepackage{bm}
\usepackage{color}

\begin{document}

\title{Singular dynamics in the failure of soft adhesive contacts}

\author{Justin D. Berman}
\affiliation{Department of Physics, Williams College, Williamstown, MA, USA}

\author{Manjari Randeria}
\affiliation{Department of Physics, Yale University, New Haven, CT, USA}%

\author{Robert W. Style}
\affiliation{Department of Materials, ETH Z\"urich, Z\"urich, Switzerland}%

\author{Qin Xu}
\affiliation{Department of Materials, ETH Z\"urich, Z\"urich, Switzerland}%

\author{James R. Nichols}
\affiliation{Department of Physics, Williams College, Williamstown, MA, USA}

\author{Aidan J. Duncan}
\affiliation{Department of Physics, Williams College, Williamstown, MA, USA}

\author{Michael Loewenberg}
\affiliation{Department of Chemical and Environmental Engineering, Yale University, New Haven, CT, USA}%

\author{Eric R. Dufresne}
\affiliation{Department of Materials, ETH Z\"urich, Z\"urich, Switzerland}%

\author{Katharine E. Jensen}
\affiliation{Department of Physics, Williams College, Williamstown, MA, USA}
 \affiliation{Department of Materials, ETH Z\"urich, Z\"urich, Switzerland}
\email{kej2@williams.edu}

\date{\today}

\begin{abstract}
We characterize the mechanical recovery of compliant silicone gels following adhesive contact failure. 
We establish broad, stable adhesive contacts between rigid microspheres and soft gels, then stretch the gels to large deformations by pulling quasi-statically on the contact. 
Eventually, the adhesive contact begins to fail, and ultimately slides to a final contact point on the bottom of the sphere.
Immediately after detachment, the gel  recoils quickly with a self-similar surface profile that evolves as a power law in time, 
suggesting that the adhesive detachment point is singular.
The singular dynamics we observe are consistent with a relaxation process driven by surface stress and slowed by viscous flow through the porous, elastic network of the gel.
Our results emphasize the importance of accounting for both the liquid and solid phases of gels in understanding their mechanics, especially under extreme deformation.
\end{abstract}

\maketitle

\section{Introduction} 

From living tissue  to food to common adhesives, 
gels are ubiquitous in nature, engineering, and industry \cite{liu1995collagen,bonnevie2015elastoviscous,abbott2003cell,doi1993gels,banerjee2012food,Creton2003}.
They are frequently used in biomedical applications 
\cite{jeong1997biodegradable,luo2004photolabile,discher2009growth,lee2012alginate} 
and present desirable opportunities to engineer medical devices with  mechanical properties similar to biological tissue \cite{noguchi1991poly,minev2015electronic}.
Recent work has explored the possibilities of gels as materials for soft robotics and other useful machines, with particular excitement around the development of super-tough hydrogels that can handle extreme mechanical manipulations without failing \cite{calvert2009hydrogels,shepherd2011multigait,sun2012highly,kim2013soft,Martinez2013}. 

Despite the many applications of these materials,  
the mechanics of gels remain a very active area of research. 
Substantial recent work has demonstrated that highly compliant materials, including soft gels, often respond to mechanical stresses very differently than their stiffer counterparts \cite{style2017elastocapillarity,andreotti2016solid}.
New physics emerges on ``elastocapillary'' length scales, comparable to the ratio of the surface stress to the elastic modulus, $\Upsilon/E$.
In this regime, which can be as large as tens of $\mu$m or even mm, soft solids begin to act like liquids that can't flow: for example, undergoing capillary instabilities  \cite{Mora2010,Chakrabarti2013,Paretkar2014,shao2018extracting}, 
forming adhesive contacts that quantitatively resemble capillary bridges  \cite{liu2016effect,xujensen2017direct,jensen2017strain,pham2017elasticity}, and blurring the line between adhesion and wetting \cite{style2013surface,Salez2013,Xu2014,cao2014elastocapillarity, henann2014modeling, jensen2015wetting,andreotti2016soft,ina2017adhesion}.
Other effects have also been found to be important for understanding gel mechanics,
including poroelasticity \cite{hui2006contact,SuoPoisson2010,hu2010using,hu2012viscoelasticity,liu2016osmocapillary,zhao2018growth}, arising from the dissipative flow of the internal fluid phase of a gel through its elastic network, and viscoelasticity, which in gels includes accounting for the power-law frequency-dependence of the storage and loss moduli typical of these materials at high frequencies \cite{shanahan1995viscoelastic,hui2016effect, karpitschka2015droplets,pandey2016lubrication,pandey2017dynamical}.

With a few notable exceptions (e.g. \cite{hui2016effect,karpitschka2015droplets,zhao2018growth}), 
most low- and high-strain studies exploring the effects of solid capillarity have focused on static or quasi-static deformations.
However, soft solids are often subjected to extreme deformations and fast dynamics, especially in the context of adhesive detachment  \cite{karpitschka2015droplets,creton2016fracture,hui2016effect}. 
Meanwhile, there exists a rich literature studying dynamic capillary instabilities and breakup in liquids 
\cite{rayleigh1878instability,keller1983surface,goldstein1993topology,shi1994cascade,zeff2000singularity,sierou2004self,eggers2008physics}, but it is not yet known whether similar singularities occur in capillary-dominated solids.

In this paper, we characterize the mechanical recovery of compliant, elastic gels after an extreme deformation and sudden release from a point contact. 
We place smooth, rigid spheres in adhesive contact with flat, sticky, silicone gels and then pull the spheres away quasi-statically until the contact line becomes unstable and slides toward a final contact point on the bottom of the sphere. 
We use high-speed video microscopy to capture the deformation of the gel surface as it recovers after detachment.
The gel  initially  recoils with a self-similar shape that evolves as approximately a 1/4 power law in time, indicating that it experienced a  singularity at the moment of detachment.
Complete recovery to a flat surface takes about 1 second.
We find that the profiles and power law exponents we observe in the self-similar regime are fundamentally different than the relaxation after breakup of either simple liquids or viscoelastic materials \cite{eggers2008physics,keller1983surface,hui2016effect}.
Instead, by applying a simple scaling argument, we show that our observations are consistent with the recoil being driven by surface stresses and slowed by 
Darcy flow 
of the gel's viscous free fluid phase through its crosslinked elastic network.
These findings emphasize the important roles played by both phases that make up a gel in determining its static and dynamic mechanical response to deformation.

\begin{figure*}[htb]
\centering
  \includegraphics[width=\textwidth]{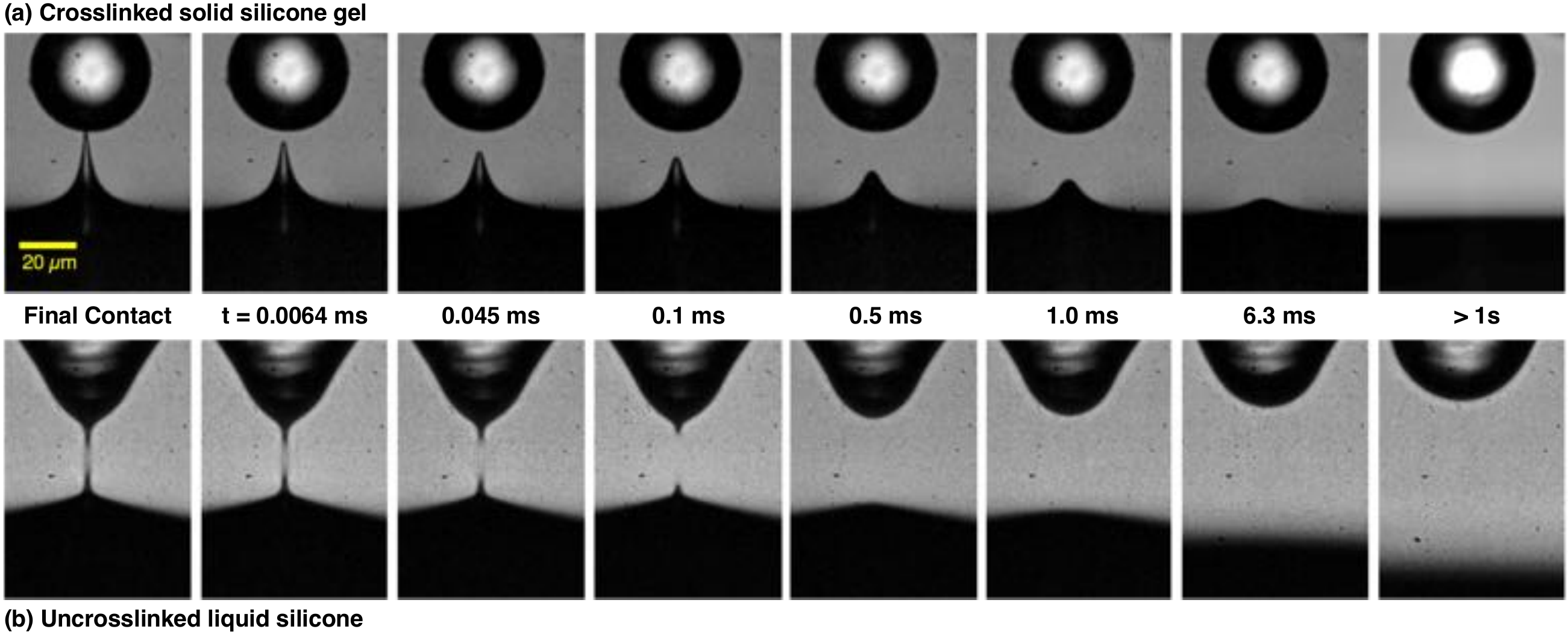}
  \caption{\label{images} Raw images of recoil after adhesive contact failure.
  (a) (left to right) A soft, silicone gel with Young's modulus $E = 5.0 \pm 0.1$ kPa detaches from a final point of contact and relaxes back to its original, flat geometry.
  (b) An uncrosslinked liquid silicone capillary bridge under the same conditions breaks contact very differently, first pinching off to a final connecting thread, then breaking and leaving fluid behind on both sides of the original contact. 
  Both  examples were filmed at 78,000 fps with a 1 $\mu$s exposure time.
}
\end{figure*}

\section{Experiment}

We prepare compliant, adhesive silicone gel substrates by mixing together liquid divinyl-terminated polydimethylsiloxane (PDMS) (Gelest, DMS-V31) with a chemical crosslinker (Gelest, HMS-301) and catalyst (Gelest, SIP6831.2).
The mixture is degassed in vacuum, set into the appropriate experimental geometry, and cured in an oven at 70$^\circ$C for at least 24 hours.
After curing, we perform all experiments at room temperature.

We control the stiffness of the silicone gels by adjusting the ratio of polymer to crosslinker.
Unless otherwise noted, the experiments described here were all performed on substrates with Young's modulus $E = 5.0 \pm 0.1$ kPa, measured by bulk indentation  using a texture analyzer (Texture Technologies, TA.XT Plus).
We also performed control experiments with stiffer ($E=8.5\pm0.1$ kPa and $E=17\pm1$ kPa) adhesive substrates, as well as with uncrosslinked liquid PDMS (Gelest DMS-V31), in order to compare the effects of varying the crosslink density or removing it entirely.
To verify that the gels do not change over time, we remeasured all moduli after storing for one month at room temperature and found no difference from the original measurements.
On the stiffer substrates, the adhesive contacts fail at smaller total displacements, but the results are otherwise similar.

The Poisson ratio of the gel elastic network is $\nu = 0.48$, previously measured using a compression test in a rheometer  \cite{jensen2015wetting,TFMpaper}.
The fraction of liquid PDMS in these gels varies with stiffness.
For the 5.0 kPa gels, the fluid fraction is about 68\% by weight, measured by solvent extraction using toluene. 
The extractable fluid fraction of the gels has the same measured viscosity as the original, uncrosslinked fluid, $\eta = 1$ Pa$\cdot$s.

To prepare substrates appropriate for imaging, we apply uncured silicone onto the millimeter-wide edge of a standard microscope slide, as in our earlier work \cite{jensen2015wetting,xujensen2017direct,jensen2017strain}.
After curing, the resulting substrate is about 300-$\mu$m-thick, flat along the length of the microscope slide, and very slightly curved (radius of curvature $\sim$700 $\mu$m) along the slide width.
Once the substrates are ready, we bring 20-to-50-$\mu$m-radius silica spheres (Polysciences, 07668) into adhesive contact.
We control the sphere position with sub-micrometer precision using a 3-axis micromanipulator stage (Narishige, MMO-203). 
The micromanipulator holds a pulled glass micropipette to which the sphere is glued using 5-minute epoxy (Elmer's) \cite{jensen2017strain}. 

As soon as the glass sphere touches the silicone substrate, the gel quickly spreads along the bottom of the sphere to establish a broad area of contact.
We allow this contact to equilibrate for about 5 minutes,
then begin to separate the sphere and substrate at a slow rate of 0.2$\mu$m/s either by moving the sphere or by moving the substrate, which is attached to the motorized microscope stage.
We pull on the adhesive contact until it becomes visibly unstable.
We then maintain the sphere at a fixed position
as the contact area slowly begins to shrink.
The contact line slides slowly at first but accelerates dramatically as it approaches the final contact point.
It usually takes between 20-60 seconds from when the contact line first destabilizes to the moment of detachment.
Example raw images from a typical solid gel detachment experiment are shown in Figure \ref{images}(a).
For comparison, we show corresponding images from a liquid detachment experiment using the uncrosslinked PDMS in Figure \ref{images}(b).

In experiments with different sized spheres, we found no effect of sphere size on the recoil dynamics other than changing the final detachment height; 
larger spheres, which establish a larger initial contact area than small spheres, are able to maintain stable adhesion to a greater displacement.
Therefore, all gel detachment results discussed here are from experiments with the same 20.8-$\mu$m-radius silica sphere.

We directly image the contact, deformation, and subsequent dynamic relaxation on an inverted optical microscope (Nikon Ti2 Eclipse).
We illuminate the sample in bright field with a low-Numerical Aperture (N.A.) condenser and image using a 40x (N.A. = 0.60) extra-long working distance objective lens.
A high-speed camera (Phantom v310) captures the recoil dynamics; we additionally use a high-resolution large field of view (FOV) camera for static imaging.
Gravity is normal to the image plane, and is negligible on these length scales.

To capture solid detachment over a large dynamic range in time, we performed repeated experiments recorded at different frame rates.
Each experiment brought the sphere into contact with a new location on the same substrate.
We acquired data at 300 frames per second (fps) with a 320-by-640$\mu$m FOV, 3200 fps with a 320-by-640$\mu$m FOV, 78,000 fps with a 64-by-128$\mu$m FOV, 220,000 fps with a 32-by-64$\mu$m FOV, and 500,000 fps with a 4-by-16 $\mu$m FOV.
All images were acquired with a 1 $\mu$s exposure time to minimize blurring.
By combining the results of the different experiments, we obtain data over five decades in time, from $\sim$1 $\mu$s to 0.1 s, and capture the entire breadth of the deformation profile at slower frame rates. 

\section{Results}

\begin{figure}[htb]
\centering
 \includegraphics[width=0.49\textwidth]{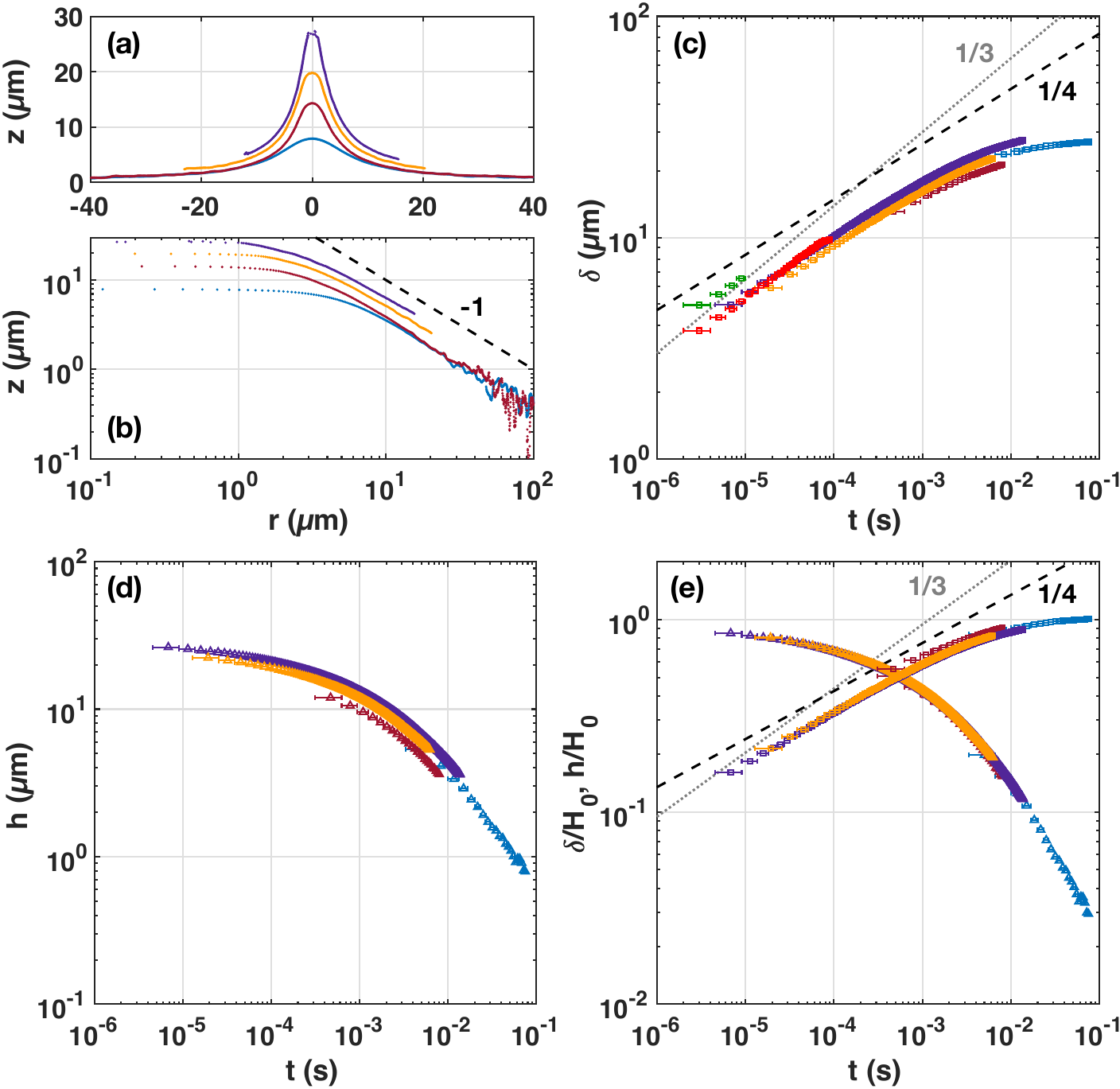}
  \caption{\label{results} Profile mapping and distance evolution.
   (a) Plots of recoiling gel surface profiles at $t=$ 2.3 $\mu$s, 71 $\mu$s, 156 $\mu$s, and 1.7 ms after detachment, mapped from detachment experiments imaged with 1 $\mu$s exposure time at 220,000 fps (purple), 78,000 fps (orange), 3200 fps (maroon), and 300 fps (blue).
   Higher frame rates capture earlier times; lower frame rates capture a larger field of view.
   (b) Log-log plot of the same data as in (a). 
  (c) Linear distance from the final contact point to the profile peak, $\delta$, versus time since detachment, $t$, for the same frame rates as above plus two experiments at 500,000 fps (red and green).
  Dashed lines have slopes of 1/3 and 1/4. 
  (d) Profile peak height, $h$, \emph{vs.} $t$.
(e) Distances normalized by initial height
, $\delta/H_0$ and $h/H_0$, versus $t$, overlaid.
Error bars are $\pm 0.5/$(frame rate).
}
\end{figure}

As we begin to pull on a stable adhesive contact,  the gel deforms with a surface profile that increasingly resembles a liquid capillary bridge  \cite{xujensen2017direct,jensen2017strain}.
However, once the contact becomes unstable, 
its approach to detachment is qualitatively different than that of a liquid, as demonstrated in Figure \ref{images}.
A liquid capillary bridge always collapses in the middle, in this case forming a final thread of fluid that subsequently breaks as the two sides recoil away from each other and leave 
fluid on both sides.
By contrast, the solid meniscus of the gel contact bridge moves up until the final contact is a single point on the bottom of the sphere, from which it detaches cleanly.
Even though a significant fraction of the gel is comprised of free fluid, 
a small amount of crosslinking is enough to completely change its behavior 
as compared to the pure fluid.

After detachment, the  gel's free surface relaxes quickly back toward its original, undeformed shape.
This motion is extremely fast at first;
based on the first-frame displacement recorded at 500,000 fps, we estimate that the tip recedes faster than 1 m/s during its initial recoil.
This is comparable to the shear wave speed of sound in the material, $c_s \approx 1.3$ m/s, but much slower than the pressure wave speed, $c_p \approx 10^3$ m/s.
The rate of relaxation slows with time, and the entire process is finished within about a second.

To quantify the recoil process, we map the gel's surface profile in each frame of each high-speed movie.
In each image frame, we locate the $(r,z)$ position of the dark edge of the gel  to about 100-nm-resolution using an adaptation of the method we developed previously,\cite{jensen2015wetting}. 
Figure \ref{results}(a-b) show example mapped profiles on both (a) linear and  (b) log-log scales from a series of experiments captured at different frame rates: 
220,000 fps, 78,000 fps, 3200 fps, and 300 fps.
Here, we have zeroed the vertical position of the profiles in reference to the final, undeformed surface long after the deformation and recovery.
The $r=0$ position is defined by the center of the axisymmetric profiles.
As evident from these plots, higher frame rates enable us to capture shorter-time dynamics, while lower frame rates capture a much larger field of view.
In Figure \ref{results}(b), we see that at large $r$ the surface displacement falls away as $1/r$, as would be expected for a purely elastic deformation of an elastic half-space \cite{hui2016effect}.
For small $r$, on the other hand, we observe that the recoiling peak rounds off rather than ending in a sharp point.
This suggests a transition from an elastically-dominated far field to a capillary-dominated near field, as we have also observed in quasi-static experiments with such soft gels  \cite{style2013surface,jensen2015wetting,xujensen2017direct,jensen2017strain}.

We characterize the peak geometry over time by fitting the data with a smoothing spline and extracting the peak center position, height $h$, and linear distance from final contact point $\delta$. 
In this way, we analyze all mapped deformation profiles except those collected at 500,000 fps.
For these, the field of view is too small to map the full surface profile, so we only obtain the linear distance from the detachment point for these data.

We plot the distance from detachment $\delta$ versus time $t$ for all experiments in Figure \ref{results}(c).
For every data set, we estimate the detachment time, $t=0$, as halfway between the last-attached and first-detached image frames, with an error of $\pm 0.5/$(frame rate).
We find that the distance from detachment grows as a power law in time of approximately $\delta \sim t^{1/4}$ over at least 2.5 decades of time, from the earliest we are able to measure (about 1 $\mu$s) until about $500\mu$s after detachment. 
At this later time, the peak has relaxed to about half of its initial height, and the recoil is just beginning to deviate from the initial power law.

We plot the peak height $h$ \emph{vs.} $t$ in Figure \ref{results}(d).
Interestingly, at late times the peak height appears to be entering a new regime of power-law decay, although the data are limited in this long-time regime.
The final detachment height varies somewhat from experiment to experiment, which slightly shifts the measured $\delta$ and $h$ values between experiments, but does not change their scaling with time, as shown in Figure \ref{results}(e).
In experiments with stiffer silicone gel substrates, we observe the same $\delta \sim t^{1/4}$ growth at early times.

\section{Discussion}

The existence of power-law evolution in length scales with time from detachment suggests that the gel is evolving from a singularity at the moment of detachment.
Singularities abound in a broad range of physical systems, including the potential energy of an electric point charge \cite{purcell2013electricity}, the rolling speed of a spinning disk as it comes to rest \cite{moffatt2000euler}, the spacetime singularity at the heart of a black hole \cite{misner2017gravitation}, and the Big Bang \cite{weinberg1972gravitation}. 
Singularities are particularly important and well-studied in a variety of fluid dynamic processes \cite{eggers2008physics}, including bursting bubbles and the eruption of liquid jets from surfaces \cite{ganan2017revision,zeff2000singularity}, the motion of an unpinned liquid contact line toward its equilibrium contact angle \cite{keller1983surface}, and the breakup and recoil of droplets and liquid capillary bridges \cite{keller1983surface,goldstein1993topology,shi1994cascade,sierou2004self,doshi2003persistence}.

Near-singularity behavior is typically characterized both by length scales that evolve as power laws in time and by self-similarity; 
without an externally-imposed length scale, the shape of a free surface will simply rescale as a power law in time with a fixed functional form \cite{barenblatt1996scaling}.
To investigate whether this occurs during recoil after soft solid adhesive detachment, we take a closer look at the post-detachment gel surface profiles in both the power-law regime and at later times.

\begin{figure}[htb]
\centering
 \includegraphics[width=0.48\textwidth]{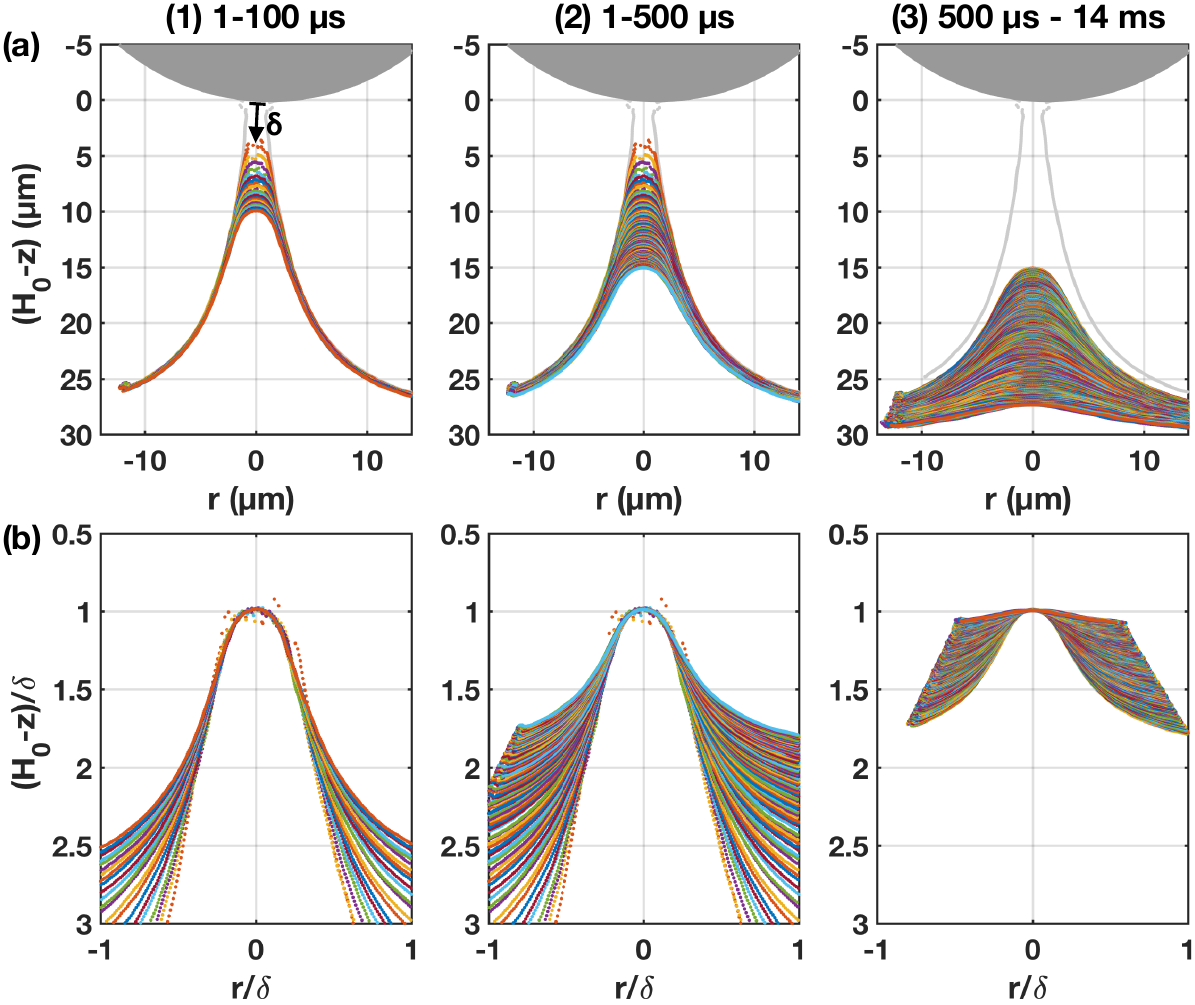}
  \caption{\label{collapse} Profiles evolving with time. 
  (a) All surface profiles from a 220,000 fps experiment.
  While the tip region recoils extremely quickly at early (1) and early-to-intermediate (2) times, the far-field surface only moves significantly at late times (3).
  (b) The same profiles rescaled by the distance from detachment, $\delta$.
  We find that all profiles collapse to a self-similar shape near the peak while $\delta$ scales as a power law in time.
  }
\end{figure}

In Figure \ref{collapse}(a), we plot  all as-mapped profiles from an experiment imaged at 220,000 fps over three time intervals:
(1)  from 1-100 $\mu$s, including early times fully within the power-law regime; 
(2) from 1-500 $\mu$s, including early-to-intermediate times as $\delta$ \emph{vs.} $t$ just begins to deviate from its initial power law;
and (3) from 500 $\mu$s through 13.6 ms, as the recoil slows further and no longer follows a power law.
Here, we put the zero of the coordinate system at the detachment point.
Examining these profiles, we see that at early times the recoiling motion is largely confined to the tip region,
while the far field surface barely moves at all.
At intermediate times, around 500 $\mu$s when the gel has recovered about halfway, we begin to see some relaxation of this far field in addition to the tip.
At longer times, the entire profile relaxes back toward its original, flat configuration.

To investigate whether the recoiling gel surface is self-similar in the tip region, we rescale the profiles with the measured distance from detachment $\delta$  and plot the dimensionless results in Figure \ref{collapse}(b).
We find that rescaling both the height and width of the profiles by $\delta$  collapses the profiles in the region close to the recoiling tip, within about $\pm \delta/4$.
At intermediate times, as the relaxation just begins to deviate from its initial power law, only the very peak remains self-similar, and at later times the profiles are no longer self-similar.

Both a qualitative characterization of the break-off process and the observed  power law growth in length scales indicate that solid gel detachment is a singular event.
In the pure fluid case, breakup and recoil is driven by surface tension and hindered by inertia, leading to length scales that grow as $t^{2/3}$ (Ref. \cite{keller1983surface,sierou2004self}).
However, for these gels, despite the demonstrated importance of solid capillarity in their quasi-static mechanical response \cite{jensen2015wetting,xujensen2017direct,jensen2017strain} as well as their large (68\%) fluid fraction, the adhesive detachment singularity belongs to a  different universality class than liquid breakup.

So what physics governs the singular dynamics of a soft gel as it recovers from an adhesive contact failure?
We can eliminate inertia, as the relaxation speed in most of this regime is much smaller than the speed of sound in the solid.
Both bulk elasticity and surface stress can dominate the mechanical response of soft solid materials \cite{style2017elastocapillarity,andreotti2016solid};
either of these could be responsible for the recoil driving force.
The dissipation mechanism could be viscoelastic; 
in particular, recent work has begun to investigate the consequences of power-law rheology on relaxation in these gels \cite{karpitschka2015droplets,pandey2016lubrication,pandey2017dynamical}.
As gels are systems with two microscopic phases, the dissipation could also be due to viscous stresses arising from relative flow of the free fluid phase of the gel through its permeable elastic network 
\cite{hu2012viscoelasticity,liu2016osmocapillary,zhao2018growth}. 
Note the qualitative distinction between viscoelasticity and viscous stresses; 
in the former, dissipation comes from local rearrangements of the elastic network, with no relative motion of the two phases that comprise the gel.
Overall, these possible driving forces and dissipation mechanisms suggest four possible balances that could govern the adhesive detachment singularity: (1) an elastic restoring stress \emph{vs.} viscoelasticity; (2) a solid surface tension restoring stress \emph{vs.} viscoelasticity; (3) elasticity \emph{vs.} viscous flow (i.e. poroelasticity); and (4) solid surface tension \emph{vs.} viscous flow.

We investigate the viscoelastic response of the gel by first measuring its rheology.
We measure the small-strain (1\%) frequency-dependence of both $G'$ and $G''$ over a range of angular frequencies from $\omega = 0.1$ s$^{-1}$ up to the rheometer's maximum angular frequency, $\omega = 200$ s$^{-1}$.
The complex rheology of gels and elastomers, including PDMS, can be described by the Chasset-Thirion model \cite{Long1996,karpitschka2015droplets,zhao2018geometrical}, with $G'(\omega) + i G''(\omega) = \mu_0(1 + (i \omega \tau)^n)$.
Fitting our rheology data to this model, we obtain $\mu_0 = 1.8$ kPa, $\tau = 0.11$ s, and $n = 0.52$.
Extrapolating to even the higher frequencies corresponding to the power-law regime of the experiment,
we expect both the storage modulus and the loss modulus scale approximately as the same power $n$ of frequency \cite{karpitschka2015droplets}:
$G' \sim G_0 \omega^n$ and $G'' \sim G_0 \omega^n$,
where $G_0$ is a constant with dimensions of [Pressure $\times$ Time$^n$].

We can now apply dimensional analysis to determine whether viscoelasticity can be the dominant dissipation mechanism in the recoiling gels.
If the relevant balance is elasticity \emph{vs.} viscoelasticity, the only parameters that the growing length scale $\delta$ can depend on are 
$t$ and $G_0$, such that $\delta = f(t,G_0)$.
However, no such scaling exists that could yield the correct dimensions for $\delta$.
Alternatively, if surface stress $\Upsilon$ is responsible for the dominant restoring force, then $\delta = f(t,\Upsilon, G_0)$.
In this case, the  only dimensionally-correct scaling is
$\delta \sim \Upsilon t^{n}/G_0 \sim t^{n} \sim t^{0.52}$,
but this exponent is inconsistent with our data.
Further, we expect the numerical value of $n$ to change with material stiffness \cite{scanlan1991composition}, but within measurement error we observe the same early-time power law scalings over more than a factor of three in modulus.
Therefore, any interpretation of the dynamics based on viscoelastic dissipation is inconsistent with our data.

Next we consider relative viscous flow between the liquid and solid components of the gel as the dominant dissipation mechanism.
In order for the stretched silicone substrate to recover its original, flat geometry, it must reincorporate all of the material that was pulled out of the bulk during the initial, slow deformation.
This requires flow of the free fluid phase through the permeable, compressible gel elastic network.

In this case, if the driving force is the elasticity of the network, then the balance between elasticity and viscous flow is poroelasticity,
which is characterised by diffusive-like motion of liquid through the gel mesh.
The fluid moves with a diffusivity $D\sim kE/\eta$, where $k$ is the gel permeability and $\eta$ is the viscosity of the free liquid \cite{hu2010using,hu2012viscoelasticity,liu2016osmocapillary,zhao2018growth}.
In poroelastic flow, we would expect $\delta\sim f(t,D) \sim \sqrt{Dt} \sim t^{1/2}$, but this scaling is again inconsistent with our data. 

By process of elimination, these scaling arguments suggest that the recoil of the stretched gel must be governed by a balance between surface stress and relative viscous flow.
A capillary driving force is further indicated because the radius of the recoiling peak is smaller than the elastocapillary length in the gel, $\Upsilon(\epsilon)/E$.
At zero strain, this length scale is around 4 $\mu$m, but can easily stretch to tens of micrometers or more for large strains $\epsilon$ \cite{xujensen2017direct,jensen2017strain}.
Hence, we expect solid capillarity to dominate locally over elasticity.

\begin{figure}[h]
\centering
\includegraphics[width=0.47\textwidth]{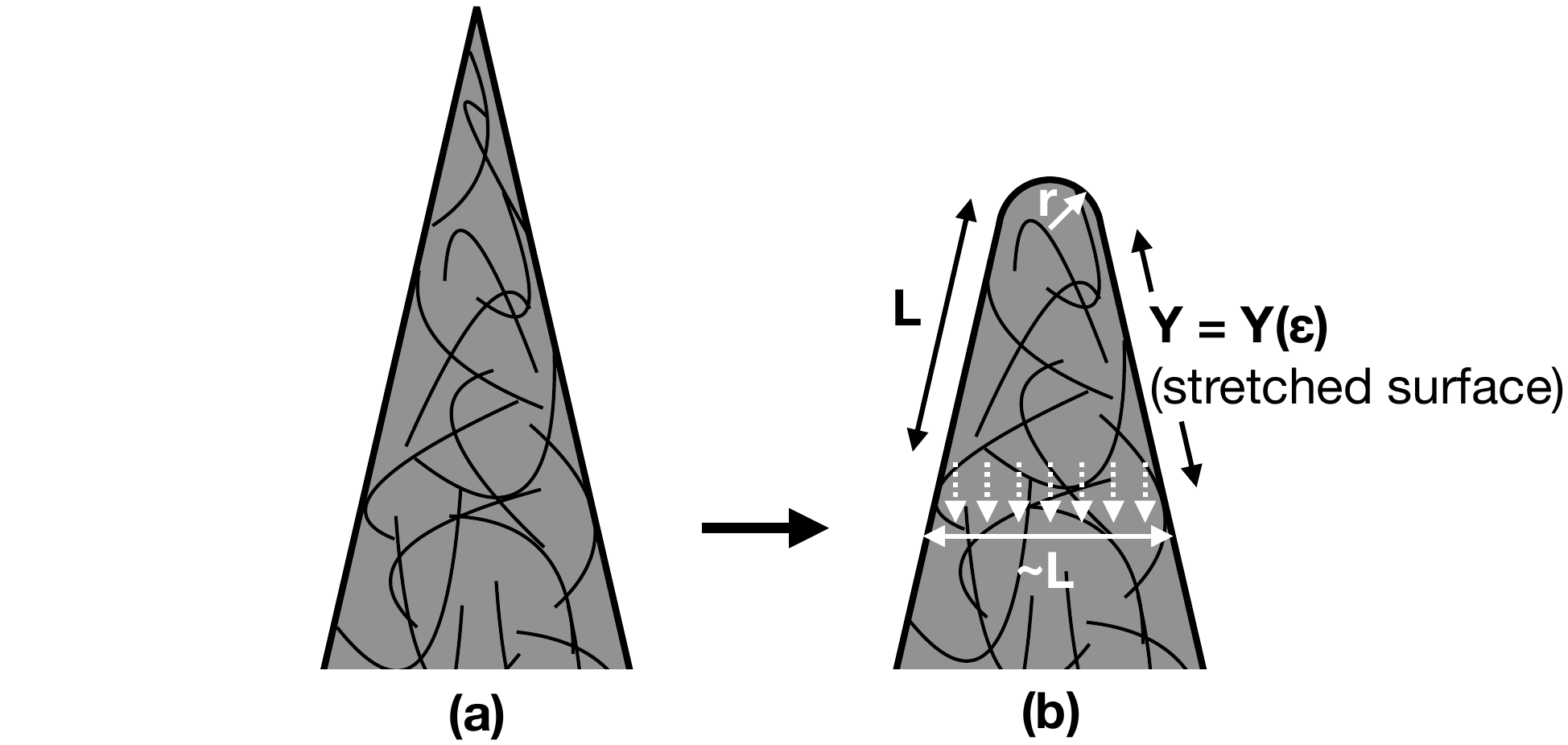}
  \caption{\label{mechanism} Schematic of proposed mechanism.
  Just prior to detachment, the gel surface resembles a cone with a sharp peak.
  Immediately after detachment, the gel assumes its similarity shape, which we approximate as having a hemispherical tip that grows in radius proportional to the distance from detachment, $r \sim \delta$.
  Solid surface stress, $\Upsilon$, creates a pressure gradient due to the surface curvature, $P \sim \Upsilon/\delta$. 
 In order to recover from this initial deformation, the internal fluid phase of the gel must flow through the gel's elastic network out of the adjacent volume, whose characteristic size we indicate as $L$, and back into the bulk of gel.
  }
\end{figure}

However, in this case, dimensional analysis of this governing balance does not yield a unique power-law solution relating $\delta$ and $t$.
Therefore, we consider a scaling argument based on the observed dynamics at the recoiling peak.
We approximate the shape of the gel surface just prior to detachment as a cone, drawn schematically in Figure \ref{mechanism}(a).
Immediately following detachment, the peak rounds off and the gel surface assumes the observed similarity solution, analogous to what occurs in liquids just after breakup \cite{keller1983surface}. 
Next, the gel surface relaxes while maintaining self-similarity at the peak throughout the  power-law regime, as sketched in Figure \ref{mechanism}(b).

The surface curvature at the peak creates a positive Laplace pressure, $P \sim \Upsilon/\delta$, which drives flow away into the bulk of the gel.
This flow can be described using Darcy's law \cite{hu2012viscoelasticity,zhao2018growth} such that the speed of this flow is given by:
\begin{equation}
v = \frac{k}{\eta} \nabla P,
\end{equation}

Since we only observe the peak and its surroundings losing volume, the draining fluid must flow into the bulk out of a region with a characteristic size $L$, as sketched in Figure \ref{mechanism}(b).
Approximating the pressure gradient as uniform over this entire region, we find that the flow velocity scales as $v\sim (k/\eta)(P/L) \sim k \Upsilon/(\eta L \delta)$.
Hence, the total volume flux out of this region scales with this flow velocity multiplied by the area ($\sim L^2$) of its base:
\begin{equation}
\dot{V}\sim \frac{k \Upsilon L}{\eta \delta}
\label{eqn:Vdot}
\end{equation}

Meanwhile, the volume lost per time from the hemispherical tip of the peak is $\dot{V} \sim d(\delta^3)/dt.$
Equating this to Equation (\ref{eqn:Vdot}), the solution of a simple differential equation yields a scaling relation between distance, time, and material properties:
\begin{equation}
\delta^4 \sim \frac{k \Upsilon L}{\eta} t
\label{scaling_result}
\end{equation}

It remains to determine the length scale $L$ that sets the size of the drainage region.
Importantly, our scaling argument makes no assumptions about $L$ other than that it sets the scale of the pressure gradient;
 in arriving at Equation \ref{scaling_result}, $L$ plays no explicit role in the singular dynamics within the similarity region.
Considering the surface geometry and gel material properties, 
two potential length scales are readily apparent: the tip radius $r$, and the elastocapillary length.
Both of these are on the order of micrometers or more, much larger than any molecular length scale.
Testing the first possibility, $L \sim r \sim \delta$, Equation \ref{scaling_result} simplifies to:
\begin{equation}
\delta \sim \left( \frac{k \Upsilon t}{\eta} \right)^{1/3} \sim t^{1/3}
\end{equation}

\noindent Due to the $\pm 0.5/$(frame rate) uncertainty in measuring the time since detachment, this scaling could be consistent with our $\delta$ \emph{vs.} $t$ data in the first decade of time that we are able to measure (from 1-10 $\mu$s), as shown in Figure \ref{results}(c).
However, it is clearly inconsistent with our measurements beyond these first few data points.

On the other hand, taking the elastocapillary length as the dominant length scale, $L \sim \Upsilon/E$, we obtain
\begin{equation}
\delta \sim \left( \frac{k \Upsilon^2 t}{\eta E} \right)^{1/4} \sim t^{1/4},
\label{yay!}
\end{equation}

\noindent which is completely consistent with our data over the entire first two and a half decades of time measured after the detachment singularity.
The established importance of elastocapillarity in static and quasi-static experiments with soft materials \cite{style2017elastocapillarity,andreotti2016solid} as well as the key role played by solid surface stress in driving the post-detachment recoil in our experiments supports our hypothesis that the size of the drainage region will be set by the elastocapillary length.
However, we note that other comparable scales for $L$ could also yield a $\delta \sim t^{1/4}$, provided that they were independent of and larger than $\delta$.
Further experiments with a wider range of substrate stiffnesses, measuring $k$ independently \cite{hu2010using}, and accounting for the dependence of $k$ on $E$ will be required to explore fully the interplay of solid surface stress and elasticity.

\section{Conclusions}
We have seen that the failure of the adhesive contact between a rigid sphere and a compliant gel is singular:
the soft gel, stretched from maintaining adhesion against a separation force, ultimately detaches from a single, final contact point.
From that moment, the deformed gel peak recoils with a self-similar shape and length scales that evolve as approximately a 1/4 power law in time. 
Although singular breakoff and recoil dynamics have been extensively studied in liquids \cite{rayleigh1878instability,keller1983surface,goldstein1993topology,shi1994cascade,zeff2000singularity,sierou2004self,eggers2008physics},  
and the gels are comprised of a significant fluid fraction,
both the surface geometry and power-law scaling of the recoiling gels demonstrate that this adhesive detachment singularity belongs to a different universality class than liquid breakup.

In exploring the physics governing the singular dynamics of the recoil process, we considered four possible balances of two driving forces, elasticity and solid surface stress, against two likely dissipation mechanisms, viscoelasticity and relative viscous flow of the gel's fluid phase through its permeable elastic network.
Dimensional scaling arguments rule out viscoelasticity as the dissipation mechanism.
A balance of elasticity and relative viscous flow, or poroelasticity, also predicts scalings inconsistent with our measurements.
Instead, we find that the post-detachment recoil dynamics are governed by a balance of solid surface stress driving against relative viscous flow of the fluid phase through the elastic network.
A scaling argument based on these physics and the observed geometry of the recoiling peak predicts power-law scalings in good agreement with what we observe in experiments.
Importantly, this provides evidence that solid capillarity plays a crucial role not only in static and quasi-static soft matter mechanics, but also in the fastest dynamical processes.

Our findings provide new insights into the dynamics of gels under extreme deformation, and emphasize the importance of accounting for both the free fluid and the solid elastic network in understanding the mechanics of gels.
Soft gels have recently emerged as a powerful testing ground for furthering our understanding of the mechanics of highly compliant materials \cite{andreotti2016solid,style2017elastocapillarity,xujensen2017direct} while simultaneously seeing increasing applications as  next-generation engineering materials in medicine,  drug delivery, microfluidics, soft robotics, and more \cite{Creton2003,jeong1997biodegradable,luo2004photolabile,discher2009growth,lee2012alginate,sokuler2009softer,noguchi1991poly,minev2015electronic,calvert2009hydrogels,shepherd2011multigait,sun2012highly,kim2013soft,Martinez2013}. 
However, much of this work has thus far treated gels as homogeneous compliant solids, neglecting their two-phase microstructure. 
Only a handful of recent studies have begun to look carefully at the mechanical response of gels in a way that takes into account their multiphase nature \cite{hu2010using,hu2012viscoelasticity,kalcioglu2012macro,jensen2015wetting,zhao2018growth}. 

In some cases, e.g. in shear deformation, we can neglect this two-phase microstructure, and work with gels as universal representatives of soft matter \cite{style2017elastocapillarity}.
However, in other cases, we will need to account carefully for the role of the internal fluid phase in determining gel structure and dynamics \cite{hu2010using,hu2012viscoelasticity,kalcioglu2012macro,jensen2015wetting,liu2016osmocapillary,zhao2018growth,pham2017elasticity,style2018contact}.
Further studies will investigate the dynamics of adhesive wet-out with soft gels as well as the approach to  detachment, both of which may also show singular dynamics.
It will also be important to understand the governing physics that can drive the free fluid phase to flow out of (or into) a soft gel, especially as this may play an important role in mediating soft contact with rough surfaces \cite{jensen2015wetting}.

We thank Frederik Brasz, Wendy Zhang, Herbert Hui, Anand Jagota, and Roderick V. Jensen for helpful discussions. 
We also thank Joey Headley and Nick Patino for help setting up the experiments.




\begin{thebibliography}{15}

\bibitem{liu1995collagen} S.~H. Liu, R.-S. Yang, R.~Al-Shaikh and J.~M. Lane, \emph{Clinical orthopaedics
  and related research}, 1995,  265--278

\bibitem[Bonnevie \emph{et~al.}(2015)Bonnevie, Galesso, Secchieri, Cohen, and
  Bonassar]{bonnevie2015elastoviscous}
E.~D. Bonnevie, D.~Galesso, C.~Secchieri, I.~Cohen and L.~J. Bonassar,
  \emph{PloS one}, 2015, \textbf{10}, e0143415. 
  
\bibitem[Abbott(2003)]{abbott2003cell}
A.~Abbott, \emph{Cell culture: biology's new dimension}, 2003. 

\bibitem[Doi(1993)]{doi1993gels}
E.~Doi, \emph{Trends in Food Science \& Technology}, 1993, \textbf{4},
  1--5. 
  
\bibitem[Banerjee and Bhattacharya(2012)]{banerjee2012food}
S.~Banerjee and S.~Bhattacharya, \emph{Critical Reviews in Food Science and
  Nutrition}, 2012, \textbf{52}, 334--346. 
  
\bibitem[Creton(2003)]{Creton2003}
C.~Creton, \emph{MRS Bull}, 2003, \textbf{28}, 434--439. 

\bibitem[Jeong \emph{et~al.}(1997)Jeong, Bae, Lee, and
  Kim]{jeong1997biodegradable}
B.~Jeong, Y.~H. Bae, D.~S. Lee and S.~W. Kim, \emph{Nature}, 1997,
  \textbf{388}, 860. 
  
\bibitem[Luo and Shoichet(2004)]{luo2004photolabile}
Y.~Luo and M.~S. Shoichet, \emph{Nature materials}, 2004, \textbf{3}, 249. 

\bibitem[Discher \emph{et~al.}(2009)Discher, Mooney, and
  Zandstra]{discher2009growth}
D.~E. Discher, D.~J. Mooney and P.~W. Zandstra, \emph{Science}, 2009,
  \textbf{324}, 1673--1677. 
  
\bibitem[Lee and Mooney(2012)]{lee2012alginate}
K.~Y. Lee and D.~J. Mooney, \emph{Progress in polymer science}, 2012,
  \textbf{37}, 106--126. 
  
\bibitem[Noguchi \emph{et~al.}(1991)Noguchi, Yamamuro, Oka, Kumar, Kotoura,
  Hyonyt, and Ikadat]{noguchi1991poly}
T.~Noguchi, T.~Yamamuro, M.~Oka, P.~Kumar, Y.~Kotoura, S.-H. Hyonyt and
  Y.~Ikadat, \emph{Journal of Applied Biomaterials}, 1991, \textbf{2},
  101--107. 
  
\bibitem[Minev \emph{et~al.}(2015)Minev, Musienko, Hirsch, Barraud, Wenger,
  Moraud, Gandar, Capogrosso, Milekovic,
  Asboth,\emph{et~al.}]{minev2015electronic}
I.~R. Minev, P.~Musienko, A.~Hirsch, Q.~Barraud, N.~Wenger, E.~M. Moraud,
  J.~Gandar, M.~Capogrosso, T.~Milekovic, L.~Asboth \emph{et~al.},
  \emph{Science}, 2015, \textbf{347}, 159--163. 
  
\bibitem[Calvert(2009)]{calvert2009hydrogels}
P.~Calvert, \emph{Advanced Materials}, 2009, \textbf{21}, 743--756. 

\bibitem[Shepherd \emph{et~al.}(2011)Shepherd, Ilievski, Choi, Morin, Stokes,
  Mazzeo, Chen, Wang, and Whitesides]{shepherd2011multigait}
R.~F. Shepherd, F.~Ilievski, W.~Choi, S.~A. Morin, A.~A. Stokes, A.~D. Mazzeo,
  X.~Chen, M.~Wang and G.~M. Whitesides, \emph{Proceedings of the national
  academy of sciences}, 2011, \textbf{108}, 20400--20403. 

\bibitem[Sun \emph{et~al.}(2012)Sun, Zhao, Illeperuma, Chaudhuri, Oh, Mooney,
  Vlassak, and Suo]{sun2012highly}
J.-Y. Sun, X.~Zhao, W.~R. Illeperuma, O.~Chaudhuri, K.~H. Oh, D.~J. Mooney,
  J.~J. Vlassak and Z.~Suo, \emph{Nature}, 2012, \textbf{489}, 133. 

\bibitem[Kim \emph{et~al.}(2013)Kim, Laschi, and Trimmer]{kim2013soft}
S.~Kim, C.~Laschi and B.~Trimmer, \emph{Trends in Biotechnology}, 2013,
  \textbf{31}, 287--294. 

\bibitem[Martinez \emph{et~al.}(2013)Martinez, Branch, Fish, Jin, Shepherd,
  Nunes, Suo, and Whitesides]{Martinez2013}
R.~V. Martinez, J.~L. Branch, C.~R. Fish, L.~Jin, R.~F. Shepherd, R.~M.~D.
  Nunes, Z.~Suo and G.~M. Whitesides, \emph{Adv Mater}, 2013, \textbf{25},
  205--212. 

\bibitem[Style \emph{et~al.}(2017)Style, Jagota, Hui, and
  Dufresne]{style2017elastocapillarity}
R.~W. Style, A.~Jagota, C.-Y. Hui and E.~R. Dufresne, \emph{Annual Review of
  Condensed Matter Physics}, 2017, \textbf{8}, 99--118. 

\bibitem[Andreotti \emph{et~al.}(2016)Andreotti, B{\"a}umchen, Boulogne,
  Daniels, Dufresne, Perrin, Salez, Snoeijer, and Style]{andreotti2016solid}
B.~Andreotti, O.~B{\"a}umchen, F.~Boulogne, K.~E. Daniels, E.~R. Dufresne,
  H.~Perrin, T.~Salez, J.~H. Snoeijer and R.~W. Style, \emph{Soft Matter},
  2016, \textbf{12}, 2993--2996. 

\bibitem[Mora \emph{et~al.}(2010)Mora, Phou, Fromental, Pismen, and
  Pomeau]{Mora2010}
S.~Mora, T.~Phou, J.-M. Fromental, L.~M. Pismen and Y.~Pomeau, \emph{Phys Rev
  Lett}, 2010, \textbf{105}, 214301. 

\bibitem[Chakrabarti and Chaudhury(2013)]{Chakrabarti2013}
A.~Chakrabarti and M.~K. Chaudhury, \emph{Langmuir}, 2013, \textbf{29},
  6926--6935. 

\bibitem[Paretkar \emph{et~al.}(2014)Paretkar, Xu, Hui, and
  Jagota]{Paretkar2014}
D.~Paretkar, X.~Xu, C.-Y. Hui and A.~Jagota, \emph{Soft Matter}, 2014,
  \textbf{10}, 4084--4090. 

\bibitem[Shao \emph{et~al.}(2018)Shao, Saylor, and
  Bostwick]{shao2018extracting}
X.~Shao, J.~Saylor and J.~B. Bostwick, \emph{Soft Matter}, 2018. 

\bibitem[Liu \emph{et~al.}(2016)Liu, Jagota, and Hui]{liu2016effect}
T.~Liu, A.~Jagota and C.-Y. Hui, \emph{Extreme Mechanics Letters}, 2016,
  \textbf{9}, 310--316. 

\bibitem[Xu \emph{et~al.}(2017)Xu, Jensen, Boltyanskiy, Sarfati, Style, and
  Dufresne]{xujensen2017direct}
Q.~Xu, K.~E. Jensen, R.~Boltyanskiy, R.~Sarfati, R.~W. Style and E.~R.
  Dufresne, \emph{Nature Communications}, 2017, \textbf{8}, 555. 

\bibitem[Jensen \emph{et~al.}(2017)Jensen, Style, Xu, and
  Dufresne]{jensen2017strain}
K.~E. Jensen, R.~W. Style, Q.~Xu and E.~R. Dufresne, \emph{Physical Review X},
  2017, \textbf{7}, 041031. 

\bibitem[Pham \emph{et~al.}(2017)Pham, Schellenberger, Kappl, and
  Butt]{pham2017elasticity}
J.~T. Pham, F.~Schellenberger, M.~Kappl and H.-J. Butt, \emph{Physical Review
  Materials}, 2017, \textbf{1}, 015602. 

\bibitem[Style \emph{et~al.}(2013)Style, Hyland, Boltyanskiy, Wettlaufer, and
  Dufresne]{style2013surface}
R.~W. Style, C.~Hyland, R.~Boltyanskiy, J.~S. Wettlaufer and E.~R. Dufresne,
  \emph{Nature Communications}, 2013, \textbf{4}, 2728. 

\bibitem[Salez \emph{et~al.}(2013)Salez, Benzaquen, and Rapha{\"e}l]{Salez2013}
T.~Salez, M.~Benzaquen and {\'E}.~Rapha{\"e}l, \emph{Soft Matter}, 2013,
  \textbf{9}, 10699--10704. 

\bibitem[Xu \emph{et~al.}(2014)Xu, Jagota, and Hui]{Xu2014}
X.~Xu, A.~Jagota and C.-Y. Hui, \emph{Soft Matter}, 2014, \textbf{10},
  4625--4632. 

\bibitem[Cao \emph{et~al.}(2014)Cao, Stevens, and
  Dobrynin]{cao2014elastocapillarity}
Z.~Cao, M.~J. Stevens and A.~V. Dobrynin, \emph{Macromolecules}, 2014,
  \textbf{47}, 6515--6521. 

\bibitem[Henann and Bertoldi(2014)]{henann2014modeling}
D.~L. Henann and K.~Bertoldi, \emph{Soft Matter}, 2014, \textbf{10},
  709--717. 

\bibitem[Jensen \emph{et~al.}(2015)Jensen, Sarfati, Style, Boltyanskiy,
  Chakrabarti, Chaudhury, and Dufresne]{jensen2015wetting}
K.~E. Jensen, R.~Sarfati, R.~W. Style, R.~Boltyanskiy, A.~Chakrabarti, M.~K.
  Chaudhury and E.~R. Dufresne, \emph{Proc Natl Acad Sci USA}, 2015,
  \textbf{112}, 14490--14494. 

\bibitem[Andreotti and Snoeijer(2016)]{andreotti2016soft}
B.~Andreotti and J.~H. Snoeijer, \emph{EPL}, 2016, \textbf{113}, 66001. 

\bibitem[Ina \emph{et~al.}(2017)Ina, Cao, Vatankhah-Varnoosfaderani, Everhart,
  Daniel, Dobrynin, and Sheiko]{ina2017adhesion}
M.~Ina, Z.~Cao, M.~Vatankhah-Varnoosfaderani, M.~H. Everhart, W.~F. Daniel,
  A.~V. Dobrynin and S.~S. Sheiko, \emph{ACS Macro Letters}, 2017, \textbf{6},
  854--858. 

\bibitem[Hui \emph{et~al.}(2006)Hui, Lin, Chuang, Shull, and
  Lin]{hui2006contact}
C.-Y. Hui, Y.~Y. Lin, F.-C. Chuang, K.~R. Shull and W.-C. Lin, \emph{Journal of
  Polymer Science Part B: Polymer Physics}, 2006, \textbf{44}, 359--370. 

\bibitem[Cai \emph{et~al.}(2010)Cai, Hu, Zhao, and Suo]{SuoPoisson2010}
S.~Cai, Y.~Hu, X.~Zhao and Z.~Suo, \emph{J Appl Phys}, 2010, \textbf{108},
  113514. 

\bibitem[Hu \emph{et~al.}(2010)Hu, Zhao, Vlassak, and Suo]{hu2010using}
Y.~Hu, X.~Zhao, J.~J. Vlassak and Z.~Suo, \emph{Applied Physics Letters}, 2010,
  \textbf{96}, 121904. 

\bibitem[Hu and Suo(2012)]{hu2012viscoelasticity}
Y.~Hu and Z.~Suo, \emph{Acta Mechanica Solida Sinica}, 2012, \textbf{25},
  441--458. 

\bibitem[Liu and Suo(2016)]{liu2016osmocapillary}
Q.~Liu and Z.~Suo, \emph{Extreme Mechanics Letters}, 2016, \textbf{7},
  27--33. 

\bibitem[Zhao \emph{et~al.}(2018)Zhao, Lequeux, Narita, Roch{\'e}, Limat, and
  Dervaux]{zhao2018growth}
M.~Zhao, F.~Lequeux, T.~Narita, M.~Roch{\'e}, L.~Limat and J.~Dervaux,
  \emph{Soft matter}, 2018, \textbf{14}, 61--72. 

\bibitem[Shanahan and Carre(1995)]{shanahan1995viscoelastic}
M.~Shanahan and A.~Carre, \emph{Langmuir}, 1995, \textbf{11}, 1396--1402. 

\bibitem[Hui and Jagota(2016)]{hui2016effect}
C.-Y. Hui and A.~Jagota, \emph{Journal of Polymer Science Part B: Polymer
  Physics}, 2016, \textbf{54}, 274--280. 

\bibitem[Karpitschka \emph{et~al.}(2015)Karpitschka, Das, van Gorcum, Perrin,
  Andreotti, and Snoeijer]{karpitschka2015droplets}
S.~Karpitschka, S.~Das, M.~van Gorcum, H.~Perrin, B.~Andreotti and J.~H.
  Snoeijer, \emph{Nature communications}, 2015, \textbf{6}, 7891. 

\bibitem[Pandey \emph{et~al.}(2016)Pandey, Karpitschka, Venner, and
  Snoeijer]{pandey2016lubrication}
A.~Pandey, S.~Karpitschka, C.~H. Venner and J.~H. Snoeijer, \emph{Journal of
  fluid mechanics}, 2016, \textbf{799}, 433--447. 

\bibitem[Pandey \emph{et~al.}(2017)Pandey, Karpitschka, Lubbers, Weijs, Botto,
  Das, Andreotti, and Snoeijer]{pandey2017dynamical}
A.~Pandey, S.~Karpitschka, L.~A. Lubbers, J.~H. Weijs, L.~Botto, S.~Das,
  B.~Andreotti and J.~H. Snoeijer, \emph{Soft matter}, 2017, \textbf{13},
  6000--6010. 

\bibitem[Creton and Ciccotti(2016)]{creton2016fracture}
C.~Creton and M.~Ciccotti, \emph{Reports on Progress in Physics}, 2016,
  \textbf{79}, 046601. 

\bibitem[Rayleigh(1878)]{rayleigh1878instability}
L.~Rayleigh, \emph{Proceedings of the London mathematical society}, 1878,
  \textbf{1}, 4--13. 

\bibitem[Keller and Miksis(1983)]{keller1983surface}
J.~B. Keller and M.~J. Miksis, \emph{SIAM Journal on Applied Mathematics},
  1983, \textbf{43}, 268--277. 

\bibitem[Goldstein \emph{et~al.}(1993)Goldstein, Pesci, and
  Shelley]{goldstein1993topology}
R.~E. Goldstein, A.~I. Pesci and M.~J. Shelley, \emph{Physical Review Letters},
  1993, \textbf{70}, 3043. 

\bibitem[Shi \emph{et~al.}(1994)Shi, Brenner, and Nagel]{shi1994cascade}
X.~Shi, M.~P. Brenner and S.~R. Nagel, \emph{Science}, 1994, \textbf{265},
  219--222. 

\bibitem[Zeff \emph{et~al.}(2000)Zeff, Kleber, Fineberg, and
  Lathrop]{zeff2000singularity}
B.~W. Zeff, B.~Kleber, J.~Fineberg and D.~P. Lathrop, \emph{Nature}, 2000,
  \textbf{403}, 401. 

\bibitem[Sierou and Lister(2004)]{sierou2004self}
A.~Sierou and J.~R. Lister, \emph{Physics of Fluids}, 2004, \textbf{16},
  1379--1394. 

\bibitem[Eggers and Villermaux(2008)]{eggers2008physics}
J.~Eggers and E.~Villermaux, \emph{Reports on progress in physics}, 2008,
  \textbf{71}, 036601. 

\bibitem[Style \emph{et~al.}(2014)Style, Boltyanskiy, German, Hyland, MacMinn,
  Mertz, Wilen, Xu, and Dufresne]{TFMpaper}
R.~W. Style, R.~Boltyanskiy, G.~K. German, C.~Hyland, C.~W. MacMinn, A.~F.
  Mertz, L.~A. Wilen, Y.~Xu and E.~R. Dufresne, \emph{Soft Matter}, 2014,
  \textbf{10}, 4047--4055. 

\bibitem[Purcell and Morin(2013)]{purcell2013electricity}
E.~M. Purcell and D.~J. Morin, \emph{Electricity and magnetism}, Cambridge
  University Press, 2013. 

\bibitem[Moffatt(2000)]{moffatt2000euler}
H.~Moffatt, \emph{Nature}, 2000, \textbf{404}, 833. 

\bibitem[Misner \emph{et~al.}(2017)Misner, Thorne, and
  Wheeler]{misner2017gravitation}
C.~W. Misner, K.~S. Thorne and J.~A. Wheeler, \emph{Gravitation}, Princeton
  University Press, 2017. 

\bibitem[Weinberg(1972)]{weinberg1972gravitation}
S.~Weinberg, \emph{Gravitation and cosmology: principles and applications of
  the general theory of relativity}, Wiley New York, 1972, vol.~1. 

\bibitem[Ga{\~n}{\'a}n-Calvo(2017)]{ganan2017revision}
A.~M. Ga{\~n}{\'a}n-Calvo, \emph{Physical review letters}, 2017, \textbf{119},
  204502. 

\bibitem[Doshi \emph{et~al.}(2003)Doshi, Cohen, Zhang, Siegel, Howell, Basaran,
  and Nagel]{doshi2003persistence}
P.~Doshi, I.~Cohen, W.~W. Zhang, M.~Siegel, P.~Howell, O.~A. Basaran and S.~R.
  Nagel, \emph{Science}, 2003, \textbf{302}, 1185--1188. 

\bibitem[Barenblatt(1996)]{barenblatt1996scaling}
G.~I. Barenblatt, \emph{Scaling, self-similarity, and intermediate asymptotics:
  dimensional analysis and intermediate asymptotics}, Cambridge University
  Press, 1996, vol.~14. 

\bibitem[Long \emph{et~al.}(1996)Long, Ajdari, and Leibler]{Long1996}
D.~Long, A.~Ajdari and L.~Leibler, \emph{Langmuir}, 1996, \textbf{12},
  5221--5230. 

\bibitem[Zhao \emph{et~al.}(2018)Zhao, Dervaux, Narita, Lequeux, Limat, and
  Roch{\'e}]{zhao2018geometrical}
M.~Zhao, J.~Dervaux, T.~Narita, F.~Lequeux, L.~Limat and M.~Roch{\'e},
  \emph{Proceedings of the National Academy of Sciences}, 2018, \textbf{115},
  1748--1753. 

\bibitem[Scanlan and Winter(1991)]{scanlan1991composition}
J.~C. Scanlan and H.~H. Winter, \emph{Macromolecules}, 1991, \textbf{24},
  47--54. 

\bibitem[Sokuler \emph{et~al.}(2009)Sokuler, Auernhammer, Roth, Liu,
  Bonacurrso, and Butt]{sokuler2009softer}
M.~Sokuler, G.~K. Auernhammer, M.~Roth, C.~Liu, E.~Bonacurrso and H.-J. Butt,
  \emph{Langmuir}, 2009, \textbf{26}, 1544--1547. 

\bibitem[Kalcioglu \emph{et~al.}(2012)Kalcioglu, Mahmoodian, Hu, Suo, and
  Van~Vliet]{kalcioglu2012macro}
Z.~I. Kalcioglu, R.~Mahmoodian, Y.~Hu, Z.~Suo and K.~J. Van~Vliet, \emph{Soft
  Matter}, 2012, \textbf{8}, 3393--3398. 

\bibitem[Style \emph{et~al.}(2018)Style, Krick, Jensen, and
  Sawyer]{style2018contact}
R.~W. Style, B.~A. Krick, K.~E. Jensen and W.~G. Sawyer, \emph{Soft Matter},
  2018, \textbf{14}, 5706--5709. 


\end{thebibliography}
%

\end{document}